\documentstyle[prb,preprint,aps]{revtex}
\draft
\input{psfig}
\psfigurepath{FIG/}

\def\AmS{{\protect\the\textfont2
        A\kern-.1667em\lower.5ex\hbox{M}\kern-.125emS}}

\makeatletter
\def\thepage{1-\@arabic\c@page}
\def\@pnumwidth{2em}
\makeatother
\begin{document}

\title{Which nanowire couples better electrically to a metal contact:
armchair or zigzag nanotube? }
\author{M. P. Anantram}
\address{NASA Ames Research Center, Mail Stop T27A-1, Moffett Field,
CA 94035-1000  }
\maketitle

\begin{abstract}
The fundamental question of how chirality affects the electronic
coupling of a nanotube to metal contacts is important for the 
application of nanotubes as nanowires. We show that metallic-zigzag
nanotubes are superior to armchair nanotubes as nanowires, by 
modeling the metal-nanotube interface. More specifically, we show that
as a function of coupling strength, the total electron transmission 
of armchair nanotubes increases and tends to be pinned close to unity
for a metal with Fermi wave vector close to that of gold. In contrast,
the total transmission of zigzag nanotubes increases to the 
maximum possible value of two. The origin of these effects lies in the
details of the wave function, which is explained.
\end{abstract}

\vspace{0.2in}

\noindent
(To appear in Applied Physics Letters, April 02, 2001 issue)

\pagebreak

A nanotube's chirality is of prime importance in determining
its electronic properties. Chirality determines whether a nanotube is
metallic or semiconducting.~\cite{Hamada92} Ref. \onlinecite{Yang99}
showed that the bandgap change with tensile and torsional strain has a
rather universal dependence on nanotube chirality. The electronic
properties of zigzag and armchair nanotubes (two distinct chiralities)
are also affected in very different manners upon 
bending.~\cite{Nardelli99}
From the view point of nanotubes in applications such as nanowires,
it is critical to understand the physics of metal-nanotube coupling.
We find that the overlap between nanotube and metal wave functions
depend significantly on chirality. As a result, metallic-zigzag 
nanotubes [which are represented by (3 times integer,0)] are superiror
to armchair nanotubes as nanowires.

We consider a single wall carbon nanotube coupled to a metal block in
the side-contacted geometry [Fig. 1]. The metal contact is treated in
the context of a free electron metal with a rectangular cross section
in the (x,z) plane, and infinite extent in the y-direction
as in most experiments. The surface Green's function of the metal
contact is calculated using standard procedures. The nanotube is 
treated using the $\pi$ orbital tight binding Hamiltonian. The coupling
between the metal and the nanotube is modeled using a tunneling-type 
Hamiltonian, which is included to all orders (and not just Born 
approximation) in calculating the transmission probability. The details
of modeling the metal-nanotube coupling can be found in reference
\onlinecite{Anantram_metal00}. The total transmission (T) is the sum
over the transmission probability of all modes at an energy.
T at energy $E$ is given by,~\cite{Datta_book}
$T(E) = Tr \left[ G^r(E) \Gamma_m (E) G^a (E) \Gamma_c (E) \right ]$,
where $\Gamma_m$ and $\Gamma_c$ are matrices that represent coupling
between the metal and a semi-infinte nanotube region either to the left
or right of the nanotube section shown in Fig. 1. $G^r$ ($G^a$) is the
full retarded (advanced) Green's function of the nanotube with coupling
to metal and semi-infinite nanotube regions included.

The coupling strength of the metal contact to the nanotube is given by
the diagonal component of $\Gamma_{m}$ which is $|t_{mc}|^2 \rho_m$,
where $\rho_m$ is the density of states of the metal surface and
$t_{mc}$ represents the hopping strength between nanotube atoms and
metal in the Hamiltonian.~\cite{Anantram_metal00} The electrical
contact length (Fig. 1) between the metal and nanotube in this work is
dictated by the available computational resources. The largest
electrical contact length considered is thirty nanotube unit
cells (approximately 72 $\AA$ and 125 $\AA$ for armchair and zigzag
nanotubes respectively).
The dimensions of the metal contact are $L_x= 400-750 \AA$ and
$L_z= 750 \AA$. 
The length of nanotube-metal electrical contact is kept constant at 
thirty nanotube unit cells, and the transmission is calculated as a 
function of coupling strength ($|t_{mc}|^2 \rho_m$).
Three values of metal Fermi wave vector ($k_f$) are considered, 1.75
$\AA^{-1}$ (Aluminum), 1.2 $\AA^{-1}$ (Gold/Silver) and 0.9 $\AA^{-1}$,
where free electron metals with $k_f$ close to the assumed values are
indicated in the parentheses. 

Fig. 2 shows the total transmission as a function of coupling
strength for a (5,5) armchair nanotube. The results show the dramatic
effect that T is pinned close to unity for $k_f= 1.2 \mbox{ and }
0.9 \AA^{-1}$.
Close to the Fermi energy (nanotube band center), two 
subbands carry current in both the positive and negative directions.
The above result indicates that only one of the two subbands couples
well to the metal. For $k_f=1.75$, T is well above unity, implying 
that both subbands couple to the metal. The wave functions of the 
crossing bands of the two positive going states of a (N,N) armchair
nanotube are:~\cite{Wallace47}
\begin{eqnarray}
\phi_{ac1} =  e^{\frac{i m_a k_a a_0}{2}} (-1)^{m_a}
                             \left[ 1 \; 1 \right]  \mbox{ and  } 
\phi_{ac2} = e^{\frac{i m_a k_a a_0}{2}}  (-1)^{m_a}
                             \left[ 1 \; -1 \right] 
                              \label{eq:acwf} \mbox{ ,}
\end{eqnarray}
where $k_a$ is the axial wave vector of the nanotube, $m_a$ is an
integer that denotes the cross section along the axial direction 
[inset of Fig. 2], and $\left[ u_1 \; u_2\right]$ is the wave
function of a unit cell of the underlying graphene sheet. For an
armchair nanotube, there is {\it no modulation} of $\left[ u_1 \;
u_2 \right]$ around the circumferential direction, for the crossing
subbands.
The wave function of one of the two subbands ($\phi_{ac2}$) is rapidly
oscillating with the nodes separated by $a_0 =1.4\AA$, in the
circumferential direction. In comparison,
the nodes of a metal wave function ($\phi_m$) with $k_f$ = 0.9, 1.2
and 1.75 $\AA^{-1}$, are separated by 6.3, 3.4, and 2 $\AA$
respectively, taking into account that the axial wave vector has to
be at least 0.75 $\AA^{-1}$.~\cite{Anantram_metal00}
As a result of this, the integral entering the Born approximation for
scattering rate, 
\begin{eqnarray}
\int \phi_{ac2}^\ast H_{c-m} \phi_m \mbox{ ,} \label{eq:born}
\end{eqnarray}
($H_{c-m}$ is the nanotube-metal coupling Hamiltonian) is very small
for $k_f = 0.9 \mbox{ and } 1.2 \AA^{-1}$, and is larger for $k_f =
1.75 \AA^{-1}$, in that order.~\cite{Datta_pc} Thus T is pinned close
to unity for $k_f = 0.9 \mbox{ and } 1.2 \AA^{-1}$, and is larger for
$k_f =1.75 \AA^{-1}$. 
Recently, Ref. \onlinecite{Choi99} discussed an alternate
mechanism by which only one of the two crossing subbands of an armchair
nanotube contributes to transport. 
The nanotube can be divided into regions where the nanotube atoms make
and do not make contact to the metal atoms.
A shift in the band structure between these two regions by about 1.5
eV causes a reflection of electrons incident from the metal into one 
of the two crossing subbands, at the interface between the two regions,
as proposed in reference \onlinecite{Choi99}.
Our work includes such a shift but in comparison to reference 
\onlinecite{Choi99}, we find that the conductance can be around unity
(for $k_f = 0.9 \mbox{ and } 1.2 \AA^{-1}$) even when this shift is 
smaller than 1.5 eV. 
Also, we propose that the
crossing subband with the smaller angular momentum contributes
more significantly to transport.

Fig. 3 shows the metal-nanotube total transmission (T) as a function
of coupling strength for a (6,0) zigzag nanotube. In stark contrast to
the armchair case, T does not saturate at unity. With increasing 
coupling strength, T approaches two, the maximum value possible. That
is, both positive 
going subbands contribute to transmission from metal to nanotube. The
wave function of the two crossing subbands of a zigzag nanotube are:
\begin{eqnarray}
\phi_{zz1} = e^{\frac{-i \sqrt{3} m_a k_a a_0}{2}} 
             e^{\frac{i 2 \pi m_a}{3}}
             e^{\frac{i 4 \pi m_c}{3}} [u_1\; u_2] 
\mbox{ and  } 
\phi_{zz2} = e^{\frac{-i \sqrt{3} m_a k_a a_0}{2}} 
             e^{\frac{i 4 \pi m_a}{3}}
             e^{\frac{i 8 \pi m_c}{3}}  [u_1\; u_2]
\label{eq:zzwf} \mbox{ ,}
\end{eqnarray}
where, $m_a$ is an integer that denotes the cross section along the 
axial direction and $m_c$ is an integer denoting the various unit
cells along the circumferential direction as shown in Fig. 3. The
wave function along the circumferential direction varies much
more slowly than the armchair wavefunction:
\begin{eqnarray}
\phi_{zz}(m_{a},m_{c}) + \phi_{zz}(m_{a},m_{c}+1) +
\phi_{zz}(m_{a},m_{c}+2) = 0 \mbox{ , } \label{eq:zz}
\end{eqnarray}
which corresponds to a distance of $3 a_0$ (7.5 $\AA$) over which the
wave function adds up to zero. As a result of this feature [Eq. 
(\ref{eq:zz})], both crossing subbands of
a zigzag nanotube couple with metals.
In Figs. 2 and 3, it is noted that for small coupling strengths, T
is larger for the armchair nanotube than the zigzag nanotube case.
This is because as a result of the small circumferential wave vector
of $\phi_{ac1}$, $\phi_{ac1}$ couples more strongly to the metal than
the sum of contributions from $\phi_{zz1}$ and $\phi_{zz2}$. It is 
pointed out that at small coupling strengths, T is
significantly larger in the case of the armchair
nanotube. This is because both crossing subbands of the zigzag
nanotube have enough angular momentum to make the overlap integral
between the metal and nanotube wave functions small. With increasing
coupling strengths, both crossing
subbands of the zigzag nanotube however eventually couple well to the
metal, unlike the armchair nanotube.

The calculations presented above consider the entire circumference of
the nanotube to be coupled to the metal contact. Such a scenario is
relevant to the experiment in Ref. \onlinecite{Frank98}, which 
resulted in a conductance of approximately $2e^2/h$. Other experiments
involve the metal making contact to only part of the circumference of
the nanotube.~\cite{Tans97} We also perform calculations corresponding
to this case. Sectors of varying
lengths are considered, and the results do not change qualitatively
from that presented below. The number of atoms around the nanotube
circumference that couple to the metal contact is shown in the legend
of Fig. 4.  The main point is that the essential features of Figs.
2 and 3 are preserved when contact is made to a sector. The
difference between a four and five atom sector is negligible in the
zigzag case. The difference between the four and five atom sectors
although small in the case of an armchair nanotubes, is larger than 
the difference for zigzag nanotubes. The reason for this, based on the
discussion of scattering rate within the Born approximation above, is
that the end odd atom [Fig. 2] corresponding to the wave function 
$\phi_{ac2}$ does not have a partner-atom to compensate (to make zero)
its contribution to the scattering rate in Eq. (\ref{eq:born}).

Two practical issues, disorder/defects and length dependence, are 
discussed next. A ten percent random variation in coupling strength 
between the nanotube atoms and the metal does not cause a significant
change in the results. From an experimental view point, a large random
variation in coupling from atom to atom in a crystalline metal is
unlikely. Defects in the nanotube such as the Stone-Wales defect will
be more effective in destroying the discussed difference.

The transmission probability of an electron from the metal to the
nanotube can be made larger either by increasing the coupling strength
or by increasing the area of electrical contact, between the nanotube
and metal. From a technological perspective, the first alternative of
small contact area (as assumed in this paper) along with strong 
coupling is more desirable. In typical experiments, the coupling 
between metal and nanotube is weak compared to the 0.2eV assumed for 
the largest coupling in Figs. 2 and 3, and the contact length is 
larger. The results of this paper are also qualitatively valid for a 
calculation where the coupling strength is constant and the electrical
contact length is increased ('coupling strength' in the x-axis of Figs.
2 - 4 should be replaced by electrical contact length).
In the case of armchair nanotubes, the state with larger angular 
momentum (which couples weakly to the metal) will eventually
contribute to conductance as the contact length is made very large.
The increase in conductance with contact length is however expected
to be slow once the state with smaller anglular momentum has coupled
to the metal.

Many factors such as the role of curvature, torsion and tension of 
armchair and zigzag nanotubes play a role in determining the 
suitablility of nanotubes as nanowires. The small curvature induced 
band gap in large diameter metallic-zigzag nanotubes predicted by
tight-binding theory is smaller than kT.~\cite{Hamada92} Further, 
reference \onlinecite{Blase94} showed that a (6,0) nanotube is a 
perfect metal, contrary to the popular belief that all small diameter
metallic-zigzag nanotubes have a small bandgap.
This lends support to the use of metallic-zigzag nanotubes as nanowires.
In this paper, we considered the role of the nanotube's electron wave
function in determining the coupling strength to a metal contact, in
the absence of significant defects.
We find that zigzag nanotubes perform better than armchair nanotubes
as nanowires. For Fermi wave vectors close to that of gold, the total
transmission (T) of side-contacted armchair tubes is pinned close to
unity.
In contrast, the total transmission in case of zigzag tubes is close 
to the maximum possible value of two. This represents a two fold 
increase in the small bias current that can be driven through a zigzag
nanoutube when compared to an armchair nanotube.

I would like to thank Supriyo Datta for useful discussions.

\pagebreak

\pagebreak

\noindent
{\bf Figure Captions:}

Fig. 1: Nanotube lying on a metal contact. The  metal contact is 
infinitely long in the y-direction (open boundaries), and thirty
unit cells of the nanotube make electrical contact to the metal.
Semi-infinite nanotube regions present to the left and right of
the nanotube section are not shown. The total-transmission (T)
is evaluated from the metal to either the semi-infinite nanotube
region to the left or right.

Fig. 2: Plot of T versus coupling strength
between metal and armchair nanotube. While for $k_f = 0.9 \mbox{ and }
1.2 \AA^{-1}$, T is pinned close to unity, for $k_f=1.75 \AA$, T is 
larger. $T_{RL}$ is the transmission from the left nanotube 
semi-infinite lead to the right nanotube semi-infinite lead, in the
presence of the metal contact inbetween the two nanotube semi-infinite
leads.

Fig. 3: Plot of T versus coupling strength
between metal and zigzag nanotube. In contrast to the armchair case,
T increases to the maximum allowed value of two with coupling strength.

Fig. 4: Plot of T versus coupling strength
between metal and nanotube for the case of a sector of the nanotube 
circumference making contact to the metal. The legend shows the number
of contiguos atoms (see inset of Fig. 2) in a unit cell making contact.
The essential features of Figs.  2 and 3 are retained. The metal Fermi
wave vector was chosen to be close to that of gold (1.2 $\AA^{-1}$).

\pagebreak

\begin{figure}[h]
\centerline{\psfig{file=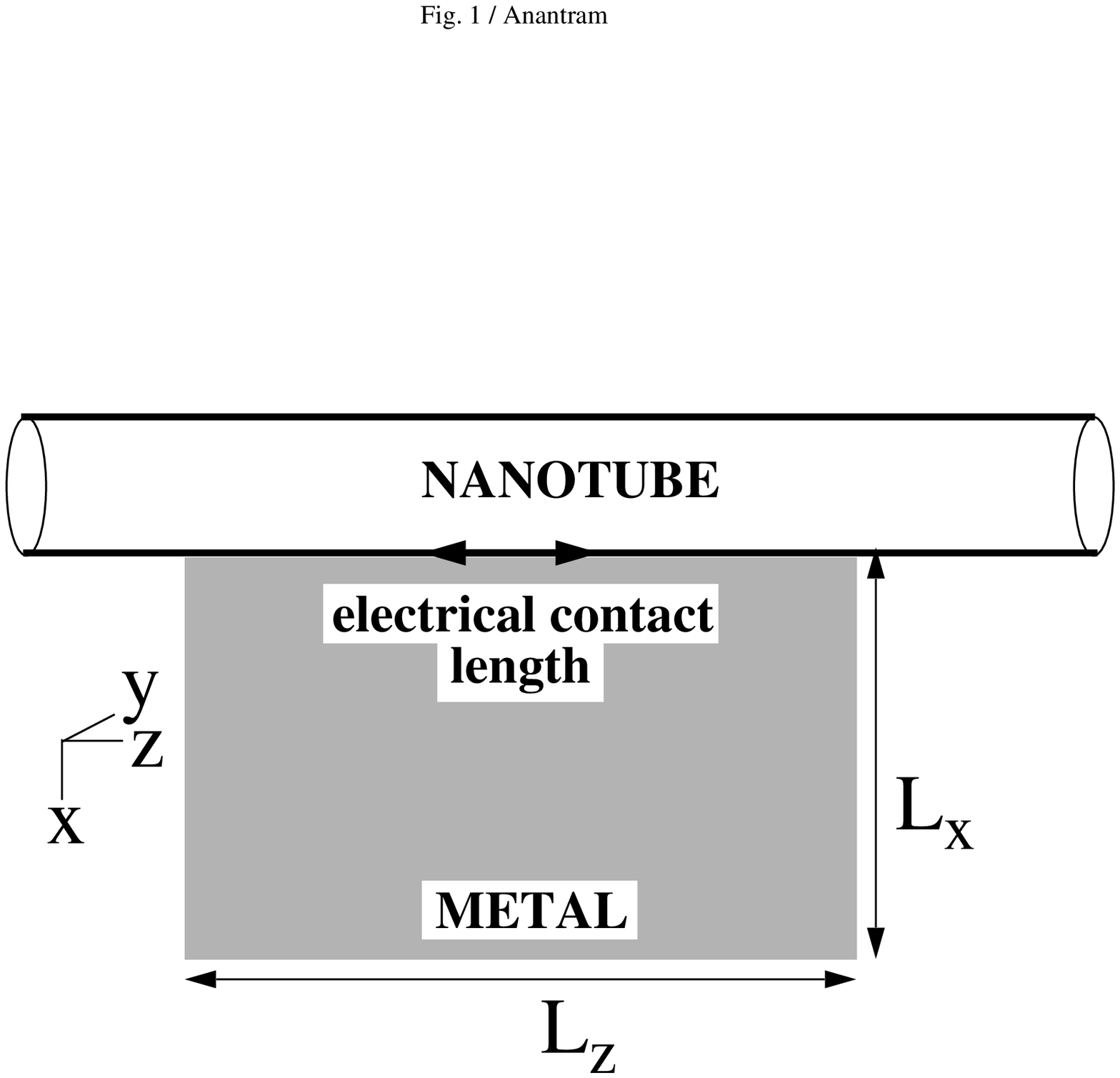 }}
\small
\end{figure}

\pagebreak

\begin{figure}[h]
\centerline{\psfig{file=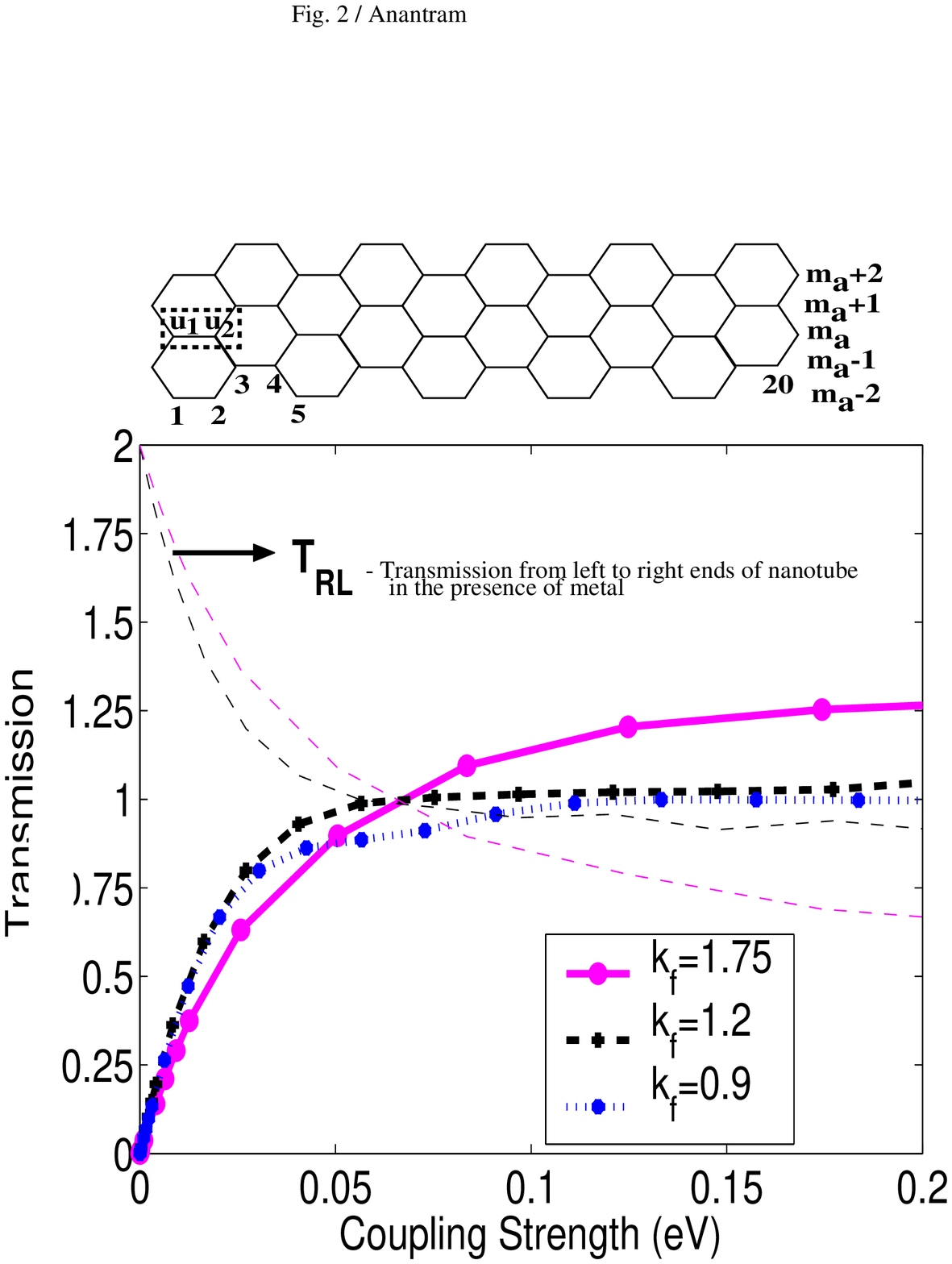}}
\small
\end{figure}

\pagebreak

\begin{figure}[h]
\centerline{\psfig{file=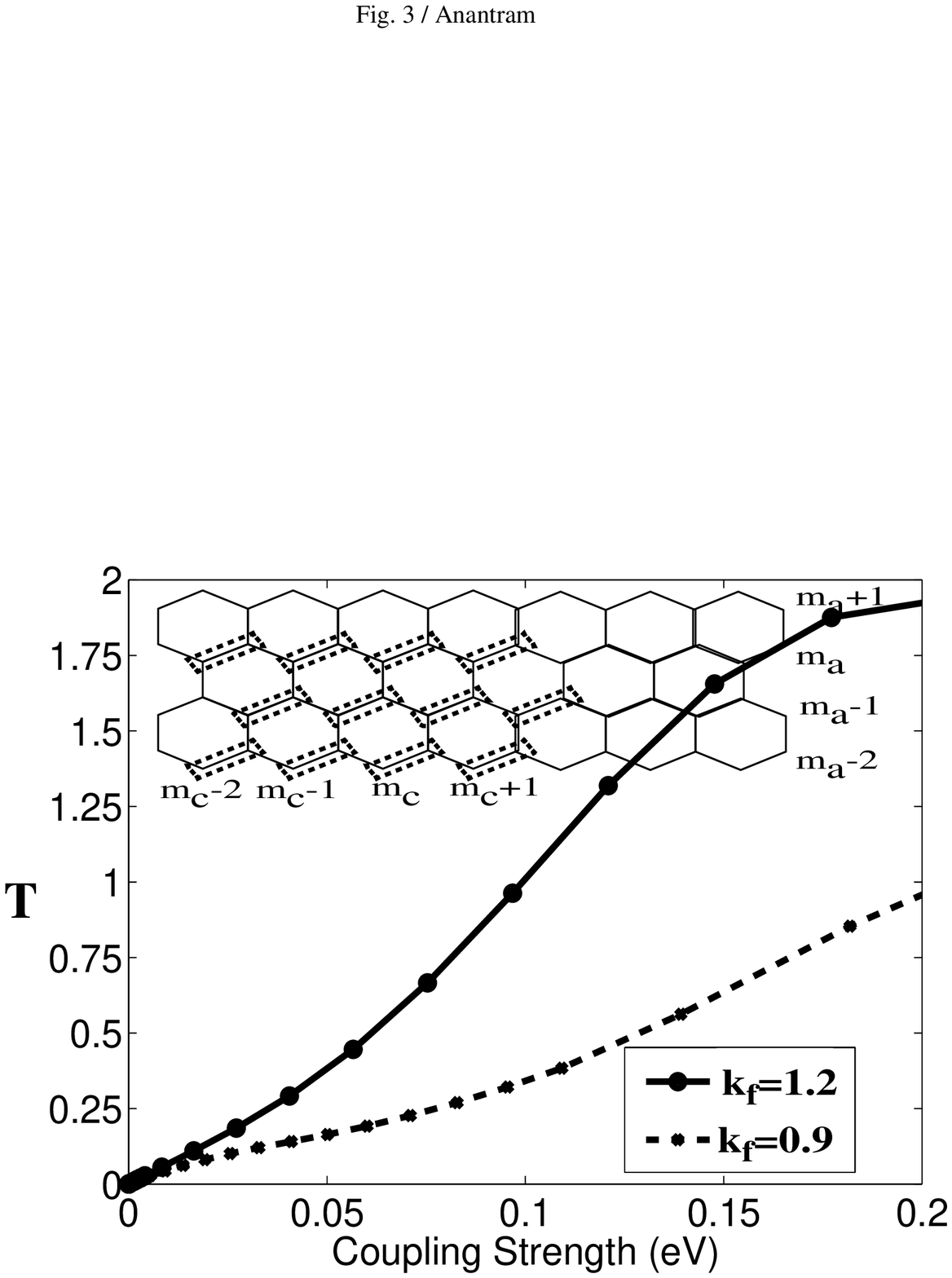}}
\small
\end{figure}

\pagebreak

\begin{figure}[h]
\centerline{\psfig{file=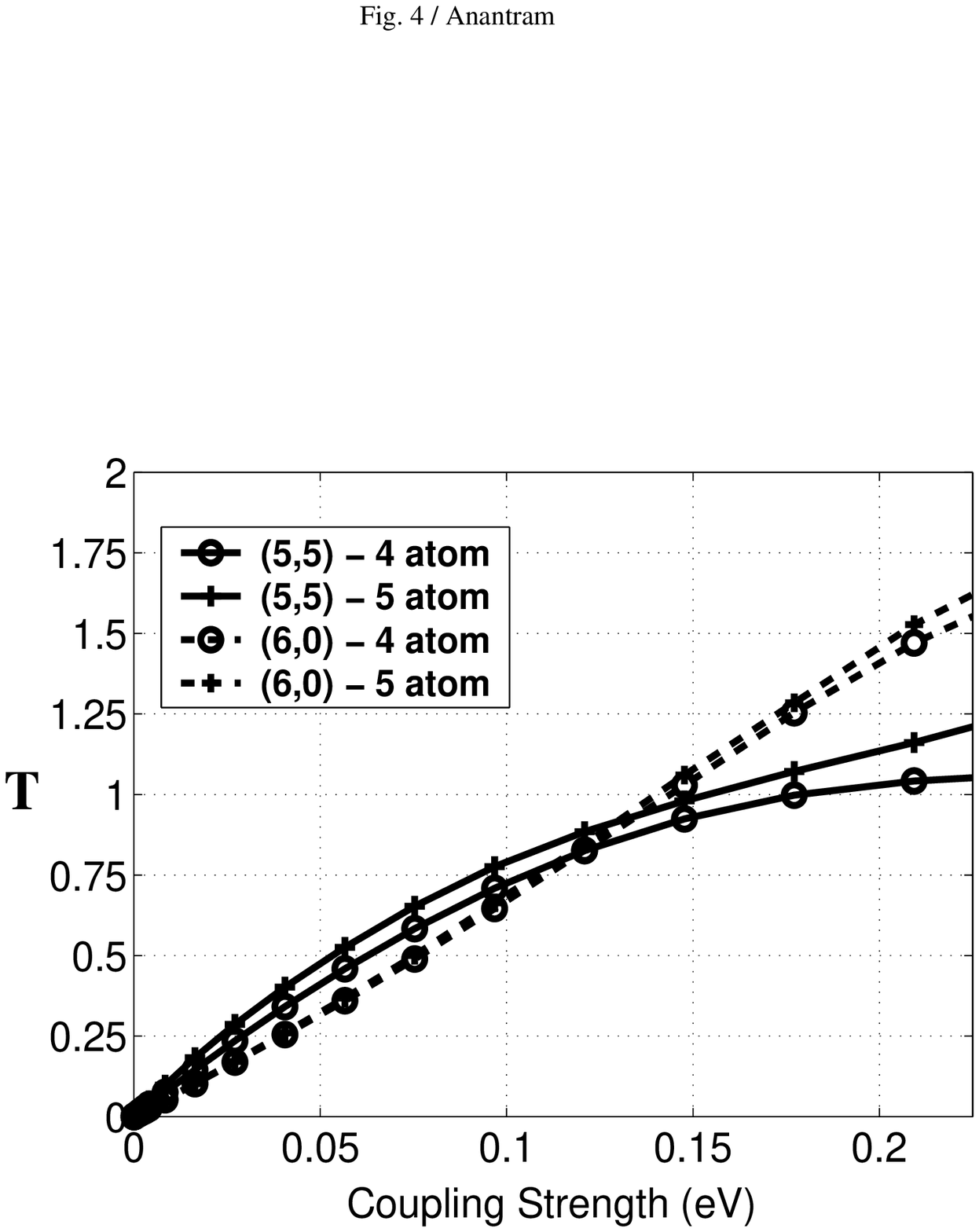}}
\small
\end{figure}

\end{document}